\documentclass[journal=jacsat,manuscript=article]{achemso}

\usepackage[version=3]{mhchem} 

\author{Shneha Biswas$^{1}$, Souren Adhikary$^{2}$ and Sudipta Dutta$^{1}$}%
 \email{sdutta@iisertirupati.ac.in}
\affiliation{
  $^{1}$Department of Physics, Indian Institute of Science Education and Research (IISER) Tirupati, Tirupati, Andhra Pradesh 517507, India.\\
  $^{2}$School of Engineering, Kwansei Gakuin University, Sanda, Hyogo 669-1330, Japan.
}

\title[An \textsf{achemso} demo]
  {Excitonic circular dichroism in boron-nitrogen clusters decorated graphene}

\begin{document}







\begin{abstract}
Within the first principle calculations, we propose a boron and nitrogen cluster incorporated graphene system for efficient valley polarization. The broken spatial inversion symmetry results in high Berry curvature at \textbf{K} and \textbf{K}$'$ valleys of the hexagonal Brillouin zone in this semiconducting system. The consideration of excitonic quasiparticles within GW approximation along with their scattering processes within many-body Bethe-Salpeter equation gives rise to an optical gap of 1.72 eV with an excitonic binding energy of 0.65 eV. Owing to the negligible intervalley scattering, the electrons in opposite valleys are selectively excited by left- and right-handed circular polarized lights, as evident from the oscillator strength calculations. Therefore, this system can exhibit circular-dichroism valley Hall effect in the presence of the in-plane electric field. Moreover, such excitonic qubits can be exploited for information processing.
\end{abstract}


Experimental demonstrations of valley-dependent optical selection rule for electronic excitations in two-dimensional (2D) semiconducting transition metal dichalcogenides (TMDs) have attracted tremendous attention in the field of valleytronics \cite{rycerz2007valley,mak2018light,mak2014valley}. This is particularly due to their applications in advanced information processing in terms of exciton qubits\cite{pacchioni2020valleytronics}. Such valleytronic property stems from the non-zero Berry curvature that acts as a pseudo-magnetic field in momentum space, arising from the broken spatial inversion symmetry ($\mathcal{P}$)\cite{vitale2018valleytronics}. This drives the electrons and holes in opposite directions in a Hall device arrangement to give rise to the circular-dichroism Hall effect under the irradiation of circularly polarized lights of a certain chirality. Owing to the opposite Berry curvatures in two time-reversal pair valleys, the directions of the charge carriers reverse when the chirality of the light is changed\cite{zeng2012valley,ye2017optical,mak2012control,cao2012valley}. Such optical selection of valley pseudo-spins in semiconducting materials can act as a unique degree of freedom in addition to the charge and spin degrees of freedom of the charge carriers.

The presence of $\mathcal{P}$ and lack of band gap at non-equivalent \textbf{K} and \textbf{K}$'$ points in monolayer graphene make the valley pseudo-spins inaccessible for optoelectronic and valleytronic applications\cite{neto2009electronic,geim2007rise}. Over the past few years, numerous efforts have been employed to break the $\mathcal{P}$ in graphene. For example, the self-assembled line defect in graphene has been demonstrated experimentally to show valley polarization\cite{gunlycke2011graphene}. The valley-selective excitation in graphene has been realized through the application of two counter-rotating circularly polarized fields\cite{chen2019circularly}. Recently, circular dichroism has been proposed in antiferromagnetic superatomic graphene lattice as well\cite{zhou2020realization}. All these require the modification of the graphene lattice by various means. Moreover, consideration of staggered sub-lattice potential, which introduces a phase difference in the wave function between opposite valleys, can show valley polarization\cite{xiao2007valley}. Such staggered sub-lattice potential can also open up a gap in the Fermi energy, as has been observed in the case of the hexagonal boron-nitride (hBN) system. However, the presence of strong intervalley scattering suppresses the valley polarization in hBN\cite{zhang2022intervalley}.
 
Lateral hybrid structures, composed of graphene and hBN represent a promising approach for achieving valley polarization\cite{song2017valley,elias2019direct,adhikary2023valley}. Such structures can be experimentally synthesized through chemical vapor deposition owing to the minimal lattice mismatch between graphene and hBN\cite{ci2010atomic}. Very recently, a few single-layer hexagonal borocarbonitride systems, consisting of periodically arranged boron (B), carbon (C), and nitrogen (N) atoms have been proposed theoretically for the realization of valley-selective circular dichroism\cite{liu2018valley,adhikary2023circular}. Recent experimental sophistication allows the realization of such systems by various means. For example, one such system, BC$_6$N has been synthesized experimentally, although as a quantum dot structure\cite{matsui2018one}. The breaking of $\mathcal{P}$ in these systems enables them to show valley selective optical excitations. 

\begin{figure}
    \centering
    \includegraphics [width=1\linewidth]{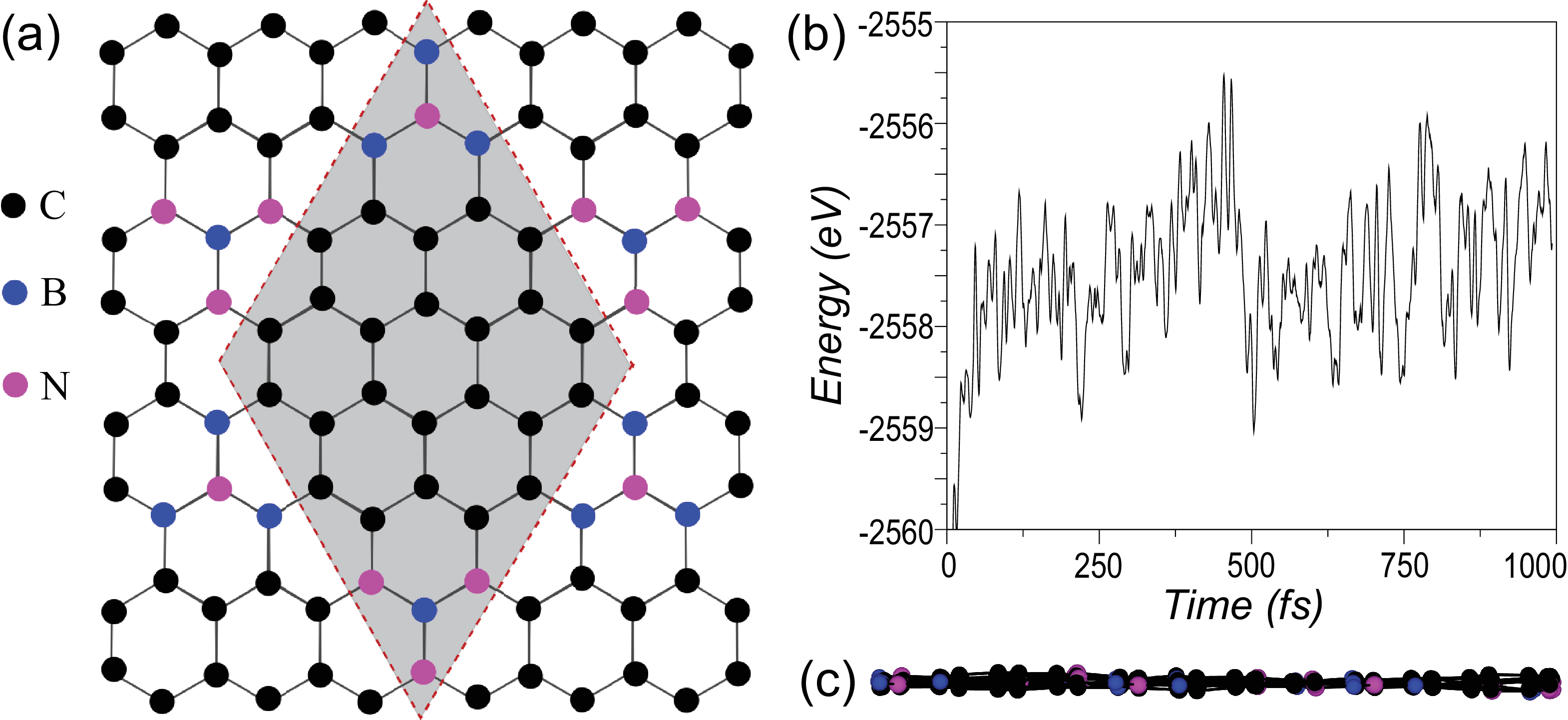}
    \caption{(a) The lattice structure of boron and nitrogen cluster incorporated graphene, named as BNcG. The shaded region represents the rhombus unit cell. (b) Total energy variation of the system as a function of time, as obtained from the AIMD calculations at 300K, considering a 3$\times$3 supercell.  (c) The side view of 3$\times$3 supercell of BNcG after 1000 fs shows the stable planner structure at room temperature.}
\end{figure}

In this paper, through density functional theory (DFT) based calculations, we propose a 2D hexagonal flat monolayer of triangular boron-nitrogen cluster substituted graphene (BNcG) system with the same stoichiometry of the BC$_6$N system, but with bigger unit cell, as depicted in figure 1 (a). The presence of complementary B and N-centered clusters in opposite ends of the rhombus unit cell breaks the $\mathcal{P}$, resulting in large Berry curvature with opposite signs at two time-reversal pair high-symmetric points, \textbf{K} and \textbf{K}$'$ in hexagonal BZ. The excitonic properties of this system have been explored within the many-body GW and Bethe-Salpeter equation (BSE), that show negligible inter valley scattering, indicating efficient valley polarization which is further confirmed through the oscillator strength estimation under circularly polarized lights of different chiralities.

\begin{figure}
    \centering
    \includegraphics [width=1\linewidth]{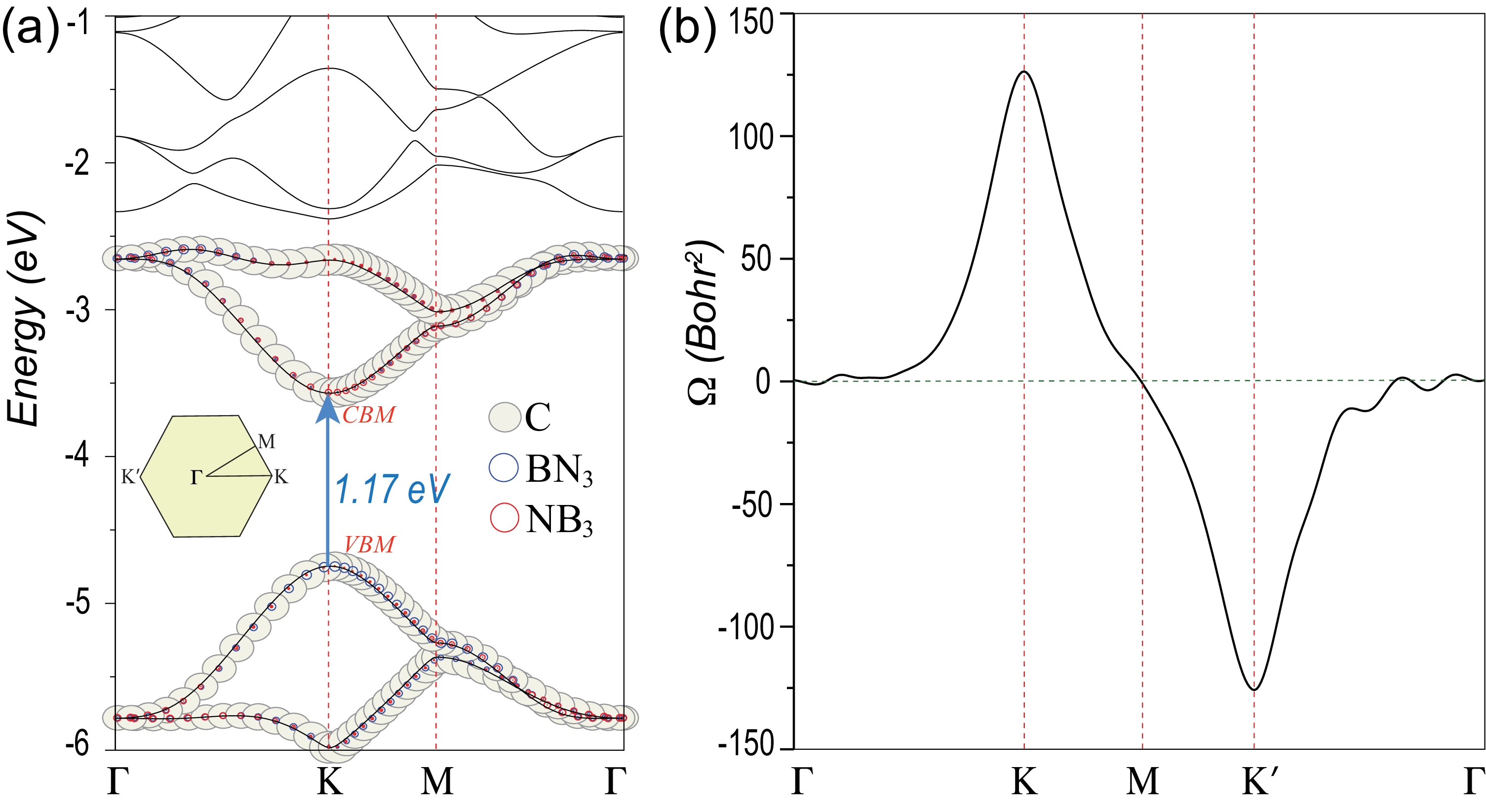}
    \caption{(a) The electronic band structure of BNcG, as obtained using PBE exchange-correlation functional. The vertical dotted lines represent the high symmetric points of the first BZ as illustrated in the inset. The direct band gap of 1.17 eV betweeen the valence band maximum (VBM) and the conduction band minimum (CBM) at \textbf{K} point is depicted by the arrow. The Fermi energy situates in between the VBM and the CBM. The contributions of all the C atoms along with BN$_3$ and NB$_3$ clusters to the frontier bands are shown by circles, where the radius of the circles denotes their extent of contributions. (b) The electronic Berry curvature of BNcG along the high symmetry path over the first BZ.}
\end{figure} 

The first-principle calculations are performed within the DFT framework using Perdew-Burke-Ernzerhof (PBE) exchange-correlation functional\cite{perdew1996generalized}, as implemented in  QUANTUM ESPRESSO package\cite{giannozzi2009quantum}. We consider norm-conserving pseudopotential available in the Pseudo-Dojo database\cite{van2018pseudodojo}. For these mean-field calculations, a plane wave basis cut-off is set to 60 Ry. The structure is fully relaxed by sampling $10\times10\times1$ Monkshort-Pack $k$-grid in the entire first Brillouin zone and the self-consistent energy convergence is achieved until the energy difference between two successive steps becomes $10^{-10}$ eV \cite{monkhorst1976special}. To avoid any interaction between the adjacent layers a vacuum space of 20\AA ~is considered in a non-periodic direction. We further calculate the Berry curvature using the Wannier90 code\cite{mostofi2008wannier90}. To investigate the structural stability in experimental conditions, we perform \textit{ab-initio} molecular dynamics (AIMD) calculation, considering NVT ensemble at 300K, as implemented in  Vienna Ab initio Simulation Package (VASP). We employ a $3\times2$ supercell, containing 192 atoms, and use the Andersen thermostat for a 2000-step simulation with a time step of 0.5 fs.

The quasi-particle (QP) band structure is then calculated within GW approximation as implemented in BerkeleyGW package\cite{deslippe2012berkeleygw}. For the GW calculation, we perform single-shot calculation for both in G (Green's function) and W (screened Coulomb potential), considering the generalized plasmon pole model. We consider $10\times10\times1$ $k$-gird sampling with the inclusion of 704 bands and 60 Ry energy cut-off to obtain the GW band structure. The QP band gap is obtained through the GW approach using Wannier90 code \cite{mostofi2008wannier90}. We check the convergence of the band gap by varying $k$-mesh, number of bands, energy cut-off, and epsilon cut-off. For optical absorption spectra calculation under plane-polarized light, parallel to the 2D plane, the BSE is solved using a fine $24\times24\times1$ coarse $k$-grid, including 4 valance and 3 conduction bands. The energy cut-off for the dielectric matrix is set to 10 Ry.

The lattice of BNcG is constructed by implanting triangular boron-nitride clusters.  Here, the hexagonal graphene domain is surrounded by alternating nitrogen-centered three boron atoms i.e., NB$_3$ clusters, and boron-centered three nitrogen atoms i.e., BN$_3$ clusters, as can be seen in Fig.1(a). The rhombus unitcell, therefore shows 25\% boron and nitrogen substitution in graphene, analogous to the 2D BC$_6$N. However, unlike the atomic substitution in the case of BC$_6$N, the BNcG system contains the periodically distributed clusters that can create a larger number of scattering centers in graphene, potentially impacting the charge carrier mobility\cite{qu2010nitrogen,usachov2011nitrogen,lherbier2008charge}. Recently, nitrogen cluster substitution in graphene has also been reported, yielding millimeter-sized single crystalline domains through an oxygen-assisted chemical vapor deposition process\cite{lin2019nitrogen}

Geometric optimization results in a planner hexagonal lattice of BNcG with lattice constant 9.9 \AA~. Note that, the ground state prefers to be nonmagnetic with no local spin preferences. To assess the stability of our proposed structure, we further perform two distinct calculations. Firstly, we calculate the cohesive energy per atom using the following formula: $E_{cohesive} = (E_{BNcG} -24 E_{C} - 3E_N -3E_B - E_{hBN})/32$, where $E_{BNcG}$ is the total energy of BNcG system containing 32 atoms, $E_C$ is the single carbon atom energy within the monolayer graphene environment, $E_N$ and $E_B$ are the energies of single nitrogen and boron atoms, respectively within sp$^2$ hybridized planar structures. Here, we consider monolayer BC$_3$ and CN$_3$ for calculating $E_B$ and $E_N$. The $E_{hBN}$ is the total energy of monolayer hBN. We consider hBN due to the presence of one boron and one nitrogen atom within both the BN$_3$ and NB$_3$ cluster regions, which essentially experience the hBN environment. We estimate the cohesive energy value as -184.4 meV per atom. The negative cohesive energy indicates the stability of the BNcG system. We further verify the thermal stability of BNcG using AIMD simulation and present the variation in total energy of the BNcG system with time in Fig.1(b). Notably, the honeycomb geometry and the planarity of the BNcG system remain unchanged (see Fig.1(c)), indicating its thermal stability even at room temperature.

Fig.2(a) depicts the electronic band structure of the BNcG system within PBE-based DFT. It shows a direct band gap of 1.17 eV at a high-symmetric point, \textbf{K}. This indicates the possibility of optical absorption in the visible spectrum. The projected band structures for the bands near the Fermi energy show a major contribution from the carbon atoms in the graphene domain. However, one can notice the minor contributions from the BN$_3$ and NB$_3$ clusters, indicating their hybridization with the graphene domain. Consequently, the characteristic Dirac bands of graphene lose the linearity and separate apart to open up a gap in Fermi energy at the corners of the hexagonal BZ. Note that, the higher band gap and its alignment with the visible spectrum make BNcG preferable over proposed superatomic graphene for semiconductor and optical applications\cite{zhou2020realization}. Moreover, the absence of vacancy patches as compared to the superatomic graphene system, provides the BNcG system higher stability that can be exploited for efficient device fabrications.

\begin{figure}
    \centering
    \includegraphics [width=1\linewidth]{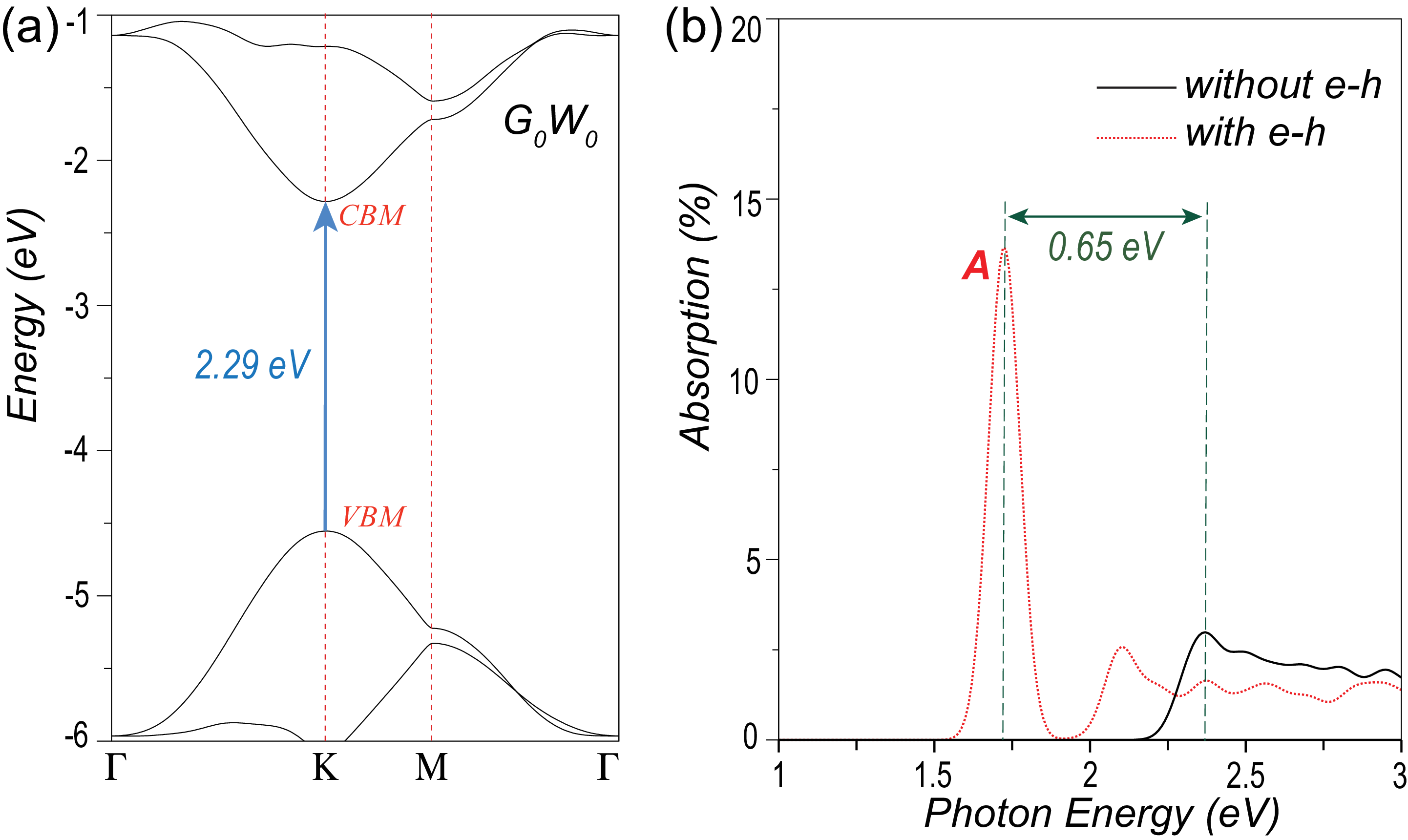}
    \caption{(a) The band structure of BNcG, as obtained from G$_{0}$W$_{0}$ calculations. The vertical dotted lines represent the high symmetric points. The direct band gap of 2.29 eV at \textbf{K} point is depicted by the arrow. (b) Optical absorption spectra of BNcG as a function of in-plane photon energy, as obtained by considering the e-h interaction (the red dotted line) and without e-h interaction (black solid line). The first singlet bright excitonic peak is marked as A. The exciton binding energy of 0.65 eV is depicted by a double-headed arrow.}
\end{figure}

To get further insight into the valleytronic properties, we calculate the Berry curvature of the BNcG system. Note that, the $\mathcal{P}$ gets broken in BNcG owing to the presence of non-equivalent BN clusters in the unit cell (see Fig.1(a)). The time-reversal symmtery however remains intact with identical energy dispersions at opposite valleys, i.e., high-symmetric points, \textbf{K} and \textbf{K}$'$. Consequently, its periodic Bl\"och wave function, $|\psi_{n\textbf{k}}\rangle$ can display varying phases considering opposite momenta. These phase-related characteristics are effectively captured by the Berry curvature. We calculate the Berry curvature of the BNcG system utilizing DFT-based Wannier wave functions, as follows:

\begin{equation}
    \Omega_{z}(\textbf{k})=-\sum_{n,m,n\neq m}\frac{2~Im \langle\psi_{n\textbf{k}}|v_{x}|\psi_{m\textbf{k}}\rangle \langle\psi_{m\textbf{k}}|v_{y}|\psi_{n\textbf{k}}\rangle}{(E_{n}-E_{m})^2}
\end{equation} 

\noindent The summation in the above equation considers all the 64 occupied bands along with 64 unoccupied bands. The term $v_{x(y)}$ is the velocity operator along the $x$ and $y$ directions, respectively. $E_{n(m)}$ is the energy eigenvalue associated with $n$-th ($m$-th) band. Fig.2(b) illustrates total Berry curvature along the high-symmetry path, revealing non-zero and opposite Berry curvatures at the \textbf{K} and \textbf{K}$'$ points. Our investigations unveil a high Berry curvature value of 125.9 Bohr$^2$ which is substantially higher than that of the other reported 2D materials\cite{zeng2012valley,ye2017optical,mak2012control,cao2012valley,zhou2020realization,adhikary2023valley}. 
Berry curvature behaves akin to a magnetic field within momentum space. Electrons possessing opposite magnetic field values at the \textbf{K} and \textbf{K}$'$ points get coupled with circularly polarized lights with opposite chirality. Therefore, by leveraging different-handed circularly polarized lights, it becomes feasible to selectively excite electrons at either the \textbf{K} or \textbf{K}$'$ valley, offering a valley-specific control in optoelectronic applications. Moreover, the Berry curvature values are opposite for the valence and conduction bands, making the holes and electrons experience opposite magnetic field. Upon application of an in-plane external electric field, these two different charge carriers at a specific valley exhibit motion in opposite directions, and this phenomenon is known as the Valley Hall effect. Changing the chirality of the circularly polarized light results in optical excitation in the other valley, and consequently, the motions of different charge carriers get reversed. Note that, the value of Berry curvature of our proposed system surpasses that of 2D BC$_6$N which stands at 96 Bohr$^2$\cite{adhikary2023circular}. This outcome arises from the substitution of BN clusters in graphene rather than isolated B or N atom substitutions, which emphasizes the greater potential of the BNcG system for valleytronics applications. 

However, having a non-zero Berry curvature is not sufficient to achieve valley polarization, as demonstrated by Yang et al. in the case of monolayer hBN system due to the reduced screened Coulomb interactions and excitonic scattering processes\cite{zhang2022intervalley}. Such intervalley excitonic scattering causes valley mixing and suppresses the valley polarization. Therefore, excitonic QP corrections within a many-body interaction picture must be considered to investigate valley-related properties accurately.

\begin{figure}
    \centering
    \includegraphics [width=1\linewidth]{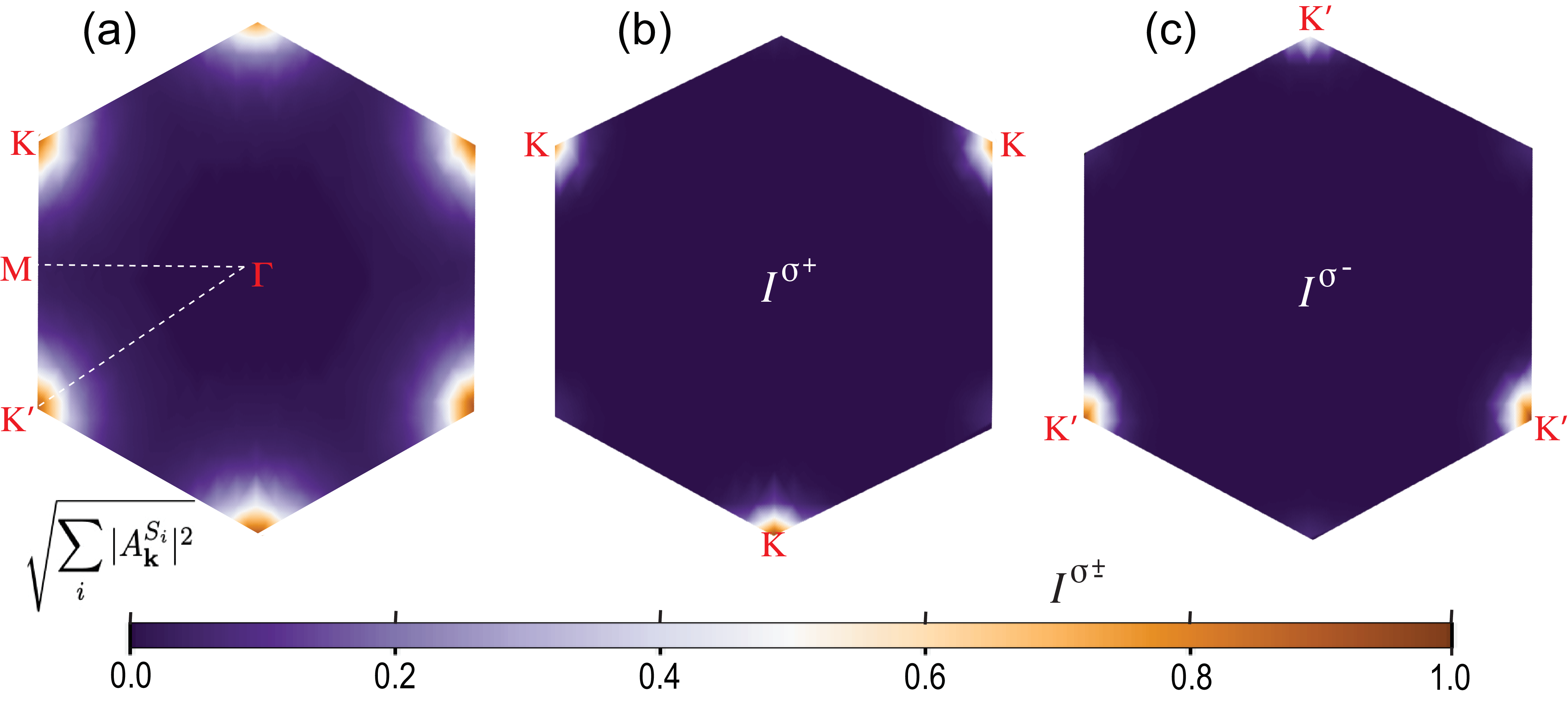}
    \caption{(a) The contour plot of the square root of the sum of the squared excitonic envelope function of the singlet bright exciton (peak A) of BNcG. The oscillator strengths of excitons in the presence of (b) left and (c) right-handed circularly polarized light show selective excitations at \textbf{K} and \textbf{K}$'$ valleys, respectively. The colour bar depicts the normalized values of the envelope function and the oscillator strength.}
\end{figure}

We calculate QP band structure by using GW approximation\cite{hybertsen1986electron} as follows:

\begin{equation} 
[-\frac{1}{2}\nabla^2 + V_{ext}+V_{H}+\Sigma(E_{nk}^{QP})]\psi_{nk}^{QP}=E_{nk}^{QP}\psi_{nk}^{Qp}
\end{equation}

\noindent where the first term is the kinetic energy of electrons or holes considering their mass and reduced Plank constant as a unit. The second term, $V_{ext}$ represents the external potential due to the ions. The third and fourth terms, $V_{H}$ and $\Sigma$ are the Hartree potential and self-energy of electrons or holes, respectively. Taking into account the converged parameters, we plot the QP band structure in Fig.3(a). The BNcG system shows a G$_0$W$_0$ gap of 2.29 eV at the high-symmetric point, \textbf{K}. This enhanced band gap as compared to the DFT gap originates from the self-energy corrections. However, the band gap still remains aligned with the visible spectrum. Note that, we further check self-consistency in self-energy corrections by one additional self-consistent update for G, i.e. G$_1$W$_0$, which shows the same band gap.

To incorporate excitonic interaction, we further solve BSE using the following equation\cite{rohlfing2000electron} 

\begin{equation}
    (E_{c\textbf{k}}^{QP}-E_{\nu\textbf{k}}^{QP})A_{\nu c\textbf{k}}^{S}+\sum_{\nu{^\prime} c^{\prime} \textbf{k}^{\prime}}\langle \nu c\textbf{k}|K^{e-h}|\nu^\prime c^\prime \textbf{k}' \rangle = \Omega^S A_{\nu c\textbf{k}}^S 
\end{equation}

\noindent where $\Omega^S$ denotes the excitonic energy, $\nu$ ($c$) represents the valence (conduction) band-index, $E_{\nu\textbf{k}}^{QP}$ ($E_{c\textbf{k}}^{QP}$) is QP energy of valence (conduction) band and $K^{e-h}$ is the electron-hole ($e-h$) interaction kernel. The term $A^S_{\nu c\textbf{k}}$ denotes excitonic envelope function. Upon obtaining the excitonic eigenvalues, we compute the absorption spectra using the imaginary part of the dielectric function

\begin{equation}
\epsilon_2 (\omega) = \frac{16\pi^2e^2}{\omega^2}\sum_S| \textbf{P}.\langle 0|\textbf{v}|S\rangle|^2 \delta(\omega-\Omega^S)
\end{equation} 

\noindent where $|0\rangle$ is the Fock space within DFT level, $|S\rangle$ is the excitonic wave function, $\omega$ is the incident photon energy, $\textbf{P}$ is the polarization vector, $\textbf{v}$ is the velocity operator, and $e$ is the electronic charge. Fig.3(b) shows the optical absorption spectra of the BNcG system without $e-h$ interaction (solid line) and with $e-h$ interaction (dotted line) under the irradiation of the linearly polarized light. The convergence of this calculation is achieved by varying the number of valance and conduction bands and with fine \textbf{k}-mesh.

In GW level (without e-h interaction) the first peak of the absorption spectra appears at 2.37 eV which shows a good agreement with the calculated GW band gap. However, within GW + BSE (with e-h interaction) level, the absorption peak (marked by A) corresponding to the first bright exciton appears at 1.72 eV which can be considered as the optical gap. The inclusion of the excitonic many-body effect through the BSE corrects the GW optical gap by 0.57 eV (2.29 eV - 1.72 eV). The first two bright exciton peaks, as can be seen in Fig.3(b), arise from doubly degenerate excitons. In between these two bright excitons, there exist two doubly degenerate dark excitons that do not contribute towards the absorption spectra. By subtracting the energy of the first bright exciton peak (A) from the energy of the first absorption peak obtained without considering e-h interaction, we estimate the excitonic binding energy as 0.65 eV, as depicted in Fig.3(b). Thus the GW + BSE absorption spectrum contains both the information of QP gap correction over the GW level theory and the formation of bound exciton which redshifts the GW + BSE absorption spectra as compared to the GW absorption spectra

Among the two degenerate excitonic states that give rise to peak A, one localizes in \textbf{K} and the other localizes in \textbf{K}$'$ valley. This degeneracy can be broken through the irradiation of circularly polarized light of different chirality\cite{adhikary2023circular,qiu2013optical}. To check the robustness of such optical selection rule for exciton formation, in Fig.4(a), we plot the \textbf{k}-resolved envelope function $\sqrt{\sum_{s_i}|A_{vc\textbf{K}}^{s_i}|^2}$ (\textit{i} denotes the degeneracy) for the first bright exciton (peak A) over the hexagonal Brillouin zone. It shows an extremely small ($3.8\%$) intervalley coupling between \textbf{K} and \textbf{K}$'$ valleys. To confirm whether this small inter-valley coupling can cause valley depolarization or not, we further calculate the excitonic oscillator strength upon irradiation of left and right-handed circularly polarized light as follows\cite{xiao2012coupled,mahon2019quantum,zhang2022intervalley}.

\begin{equation}
I^{\sigma^{\pm}}=\sum_{s_i}|A_{vc\textbf{k}}^{s_i}.\textbf{P}.\langle v\textbf{k}|\hat{\textbf{v}}|c\textbf{k}\rangle|^2
\end{equation}

\noindent where the term $\langle v\textbf{k}|\hat{\textbf{v}}|c\textbf{k}\rangle$ denotes the optical matrix element, that couples with left and right-handed circularly polarized lights, $\sigma^{\pm}$ with polarization vector \textbf{P} at opposite valleys. We plot the oscillator strength for left-($\sigma^{+}$) and right-handed ($\sigma^{-}$) circularly polarized lights in the first BZ as in Fig.4(b) and (c) for the first bright exciton (peak A). As can be seen, the $\sigma^{+}$ and $\sigma^{-}$ lights selectively excites the \textbf{K} and \textbf{K}$'$ valleys. This clearly indicates selective excitation of only one valley with a certain helicity of circularly polarized light. Therefore, the absence of intervalley scattering along with high Berry curvature results in efficient valley polarization and circular dichroism Hall effect in the BNcG system in the presence of the in-plane electric field. 
 
In conclusion, we propose a 2D honeycomb superlattice of graphene with periodic incorporation of B and N clusters with overall BC$_6$N stoichiometry. The cohesive energy calculation along with the AIMD simulations proves the stability of this system even at room temperature. The presence of B and N clusters breaks the spatial inversion symmetry, however, preserving the time-reversal symmetry with a semiconducting gap at \textbf{K} and \textbf{K}$'$ valleys. Owing to that, the system exhibits a very high Berry curvature of 125.9 Bohr$^2$ at opposite valleys, which is considerably higher than that of other reported 2D materials. Explicit consideration of excitonic quasiparticles along with their scattering possibilities within GW and many-body BSE formalism reveals an optical gap of 1.72 eV and a strong excitonic binding energy of 0.65 eV. Negligible intervalley scattering between the opposite valleys, as indicated by the envelope function calculation, ensures the robustness of the optical selection rule for valley selective optical excitations. This has been further established by explicit oscillator strength estimations in the presence of left and right-handed circular polarized lights. Such efficient valley polarization can be exploited for circular dichroism valley Hall effect and excitonic qubit applications in advanced information processing.

\begin{acknowledgement}
SB, SA and SD thank IISER Tirupati for Intramural Funding and Science and Engineering Research Board, Dept. of Science and Technology, Govt. of India for research grant (CRG/2021/001731). The authors acknowledge National Supercomputing Mission (NSM) for providing computing resources of ‘PARAM Brahma’ at IISER Pune, which is implemented by C-DAC and supported by the Ministry of Electronics and Information Technology (MeitY) and DST, Govt. of India.
\end{acknowledgement}


\bibliography{acs-achemso}

\end{document}